\documentclass [12pt] {article}
\usepackage{a4}

  \def\pp{{\mathchoice
          {%displaystyle
              \kern 1pt%
              \raise 1pt
              \vbox{\hrule width5pt height0.4pt depth0pt
                    \kern -2pt
                    \hbox{\kern 2.3pt
                          \vrule width0.4pt height6pt depth0pt
                          }
                    \kern -2pt
                    \hrule width5pt height0.4pt depth0pt}%
                    \kern 1pt
           }
            {%textstyle
              \kern 1pt%
              \raise 1pt
              \vbox{\hrule width4.3pt height0.4pt depth0pt
                    \kern -1.8pt
                    \hbox{\kern 1.95pt
                          \vrule width0.4pt height5.4pt depth0pt
                          }
                    \kern -1.8pt
                    \hrule width4.3pt height0.4pt depth0pt}%
                    \kern 1pt
            }
            {%scriptstyle
              \kern 0.5pt%
              \raise 1pt
              \vbox{\hrule width4.0pt height0.3pt depth0pt
                    \kern -1.9pt  %[e]=0.15pt
                    \hbox{\kern 1.85pt
                          \vrule width0.3pt height5.7pt depth0pt
                          }
                    \kern -1.9pt
                    \hrule width4.0pt height0.3pt depth0pt}%
                    \kern 0.5pt
            }
            {%scriptscriptstyle
              \kern 0.5pt%
              \raise 1pt
              \vbox{\hrule width3.6pt height0.3pt depth0pt
                    \kern -1.5pt
                    \hbox{\kern 1.65pt
                          \vrule width0.3pt height4.5pt depth0pt
                          }
                    \kern -1.5pt
                    \hrule width3.6pt height0.3pt depth0pt}%
                    \kern 0.5pt%}
            }
        }}

  \def\mm{{\mathchoice
                       {%displaystyle
                             \kern 1pt
               \raise 1pt    \vbox{\hrule width5pt height0.4pt depth0pt
                                  \kern 2pt
                                  \hrule width5pt height0.4pt depth0pt}
                             \kern 1pt}
                       {%textstyle
                            \kern 1pt
               \raise 1pt \vbox{\hrule width4.3pt height0.4pt depth0pt
                                  \kern 1.8pt
                                  \hrule width4.3pt height0.4pt depth0pt}
                             \kern 1pt}
                       {%scriptstyle
                            \kern 0.5pt
               \raise 1pt
                            \vbox{\hrule width4.0pt height0.3pt depth0pt
                                  \kern 1.9pt
                                  \hrule width4.0pt height0.3pt depth0pt}
                            \kern 1pt}
                       {%scriptscriptstyle
                           \kern 0.5pt
             \raise 1pt  \vbox{\hrule width3.6pt height0.3pt depth0pt
                                  \kern 1.5pt
                                  \hrule width3.6pt height0.3pt depth0pt}
                           \kern 0.5pt}
                       }}

\catcode`@=11
\def\un#1{\relax\ifmmode\@@underline#1\else
        $\@@underline{\hbox{#1}}$\relax\fi}
\catcode`@=12

\let\du=\du                     % dot-under
\def\a{\alpha}
\def\b{\beta}

\def\d{\delta}

\def\f{\phi}
\def\g{\gamma}
\def\h{\eta}

\def\j{\psi}
\def\k{\kappa}
\def\l{\lambda}

\def\o{\omega}
\def\p{\pi}
\def\q{\theta}
\def\r{\rho}
\def\s{\sigma}

\def\x{\xi}

\def\F{\Phi}

\def\J{\Psi}
\def\L{\Lambda}
\def\O{\Omega}

\def\Q{\Theta}

\def\ve{\varepsilon}

\def\cd{{\cal D}}

\def\cy{{\cal Y}}

\def\bo{{\raise-.5ex\hbox{\large$\Box$}}}               % D'Alembertian
\def\pa{\partial}                                       % curly d
\def\TH{{\raise.2ex\hbox{$\displaystyle \bigodot$}\mskip-4.7mu \llap H \;}}
\def\face{{\raise.2ex\hbox{$\displaystyle \bigodot$}\mskip-2.2mu \llap {$\ddot
        \smile$}}}                                      % happy face
\def\leftrightarrowfill{$\mathsurround=0pt \mathord\leftarrow \mkern-6mu
        \cleaders\hbox{$\mkern-2mu \mathord- \mkern-2mu$}\hfill
        \mkern-6mu \mathord\rightarrow$}
\def\dvec#1{\vbox{\ialign{##\crcr
        \leftrightarrowfill\crcr\noalign{\kern-1pt\nointerlineskip}
        $\hfil\displaystyle{#1}\hfil$\crcr}}}           % <--> accent
\def\dt#1{{\buildrel {\hbox{\LARGE .}} \over {#1}}}     % dot-over for sp/sb
\def\frac#1#2{{\textstyle{#1\over\vphantom2\smash{\raise.20ex
        \hbox{$\scriptstyle{#2}$}}}}}                   % fraction
\def\sfrac#1#2{{\vphantom1\smash{\lower.5ex\hbox{\small$#1$}}\over
        \vphantom1\smash{\raise.4ex\hbox{\small$#2$}}}} % alternate fraction
\def\bfrac#1#2{{\vphantom1\smash{\lower.5ex\hbox{$#1$}}\over
        \vphantom1\smash{\raise.3ex\hbox{$#2$}}}}       % "
\def\afrac#1#2{{\vphantom1\smash{\lower.5ex\hbox{$#1$}}\over#2}}    % "
\def\[{\lfloor{\hskip 0.35pt}\!\!\!\lceil}
\def\]{\rfloor{\hskip 0.35pt}\!\!\!\rceil}
\def\Lag{{\cal L}}
\def\du#1#2{_{#1}{}^{#2}}

\def\fracm#1#2{\hbox{\large{${\frac{{#1}}{{#2}}}$}}}
\def\ha{{\fracmm12}}

\def\un{\underline}
\def\fracmm#1#2{{{#1}\over{#2}}}

\def\low#1{{\raise -3pt\hbox{${\hskip 0.75pt}\!_{#1}$}}}

\def\Dot#1{\buildrel{_{_{\hskip 0.01in}\bullet}}\over{#1}}
\def\dt#1{\Dot{#1}}

\newskip\humongous \humongous=0pt plus 1000pt minus 1000pt
\def\caja{\mathsurround=0pt}
\def\eqalign#1{\,\vcenter{\openup2\jot \caja
        \ialign{\strut \hfil$\displaystyle{##}$&$
        \displaystyle{{}##}$\hfil\crcr#1\crcr}}\,}
\newif\ifdtup

\parskip=\medskipamount                 % space between paragraphs (LaTeX)
\thicklines                         % thick straight lines for pictures (LaTeX)

\begin{document}
\thispagestyle{empty}

{\hbox to\hsize{
\vbox{\noindent June 2003 \hfill hep-th/0304002 }}}

{\hbox to\hsize{
\vbox{\noindent revised version }}}

\noindent
\vskip1.3cm
\begin{center}

{\Large\bf    More on the Gauge-Fixed D3-Brane Action with 
\vglue.1in     Dilaton-Axion Coupling from N=1 Superspace~\footnote{
Supported in part by the `Deutsche Forschungsgemeinschaft'
 and the `Volkswagen Stiftung'}}

\vglue.2in

                  Sergei V. Ketov
\vglue.1in
     {\it Institut f\"ur Theoretische Physik\\
             Universit\"at Hannover\\
          Appelstr. 2, 30167 Hannover, Germany}\\
                         and
\\
 {\it          Department of Physics\\
          Tokyo Metropolitan University\\
         Hachioji-shi, Tokyo 192--0397, Japan}
\vglue.1in
{\sl ketov@itp.uni-hannover.de, ketov@phys.metro-u.ac.jp}

\end{center}

\vglue.3in

\begin{center}
{\Large\bf Abstract}
\end{center}

\noindent
The gauge-fixed action of a `spacetime-filling' D3-brane with dilaton-axion
coupling is formulated in N=1 superspace. We investigate its symmetries by 
paying special attention to a possible non-linearly realized extra 
supersymmetry, and emphasize the need of a linear superfield coupled to 
an abelian Chern-Simons superfield to represent a dilaton-axion 
supermultiplet in the off-shell manifestly supersymmetric approach.

\newpage

\section{Introduction}

The supersymmetric D-brane actions with local fermionic kappa symmetry were
constructed in ref.~\cite{one}. When the kappa-symmetry is fixed, half of 
supersymmetry is spontaneously broken, whereas the fermionic superpartner (with
respect to unbroken half of supersymmetry) of the $U(1)$ gauge field in the 
D-brane worldvolume can be identified with the Goldstone fermion. The most 
relevant part of the gauge-fixed D-brane action is given by a supersymmetric
{\it Born-Infeld} (BI) action \cite{one}. Gauge-fixing results in the D-brane 
actions whose all supersymmetries are non-linearly realized, i.e. non-manifest.
Unbroken supersymmetries can sometimes be made manifest by using superspace 
\cite{two,two1}.

The electric-magnetic self-duality of the BI action can be extended to a full 
$SL(2,{\bf Z})$ duality in the case of a gauge-fixed `spacetime-filling' 
D3-brane with axion-dilaton coupling \cite{three}. This feature can be 
made manifest when considering the D3-brane action as the double dimensionally
 reduced M5-brane action on a 2-torus \cite{four}. The dilaton-axion can be 
identified with the complex structure of the torus, while the $SL(2,{\bf Z})$ 
self-duality of a D3-brane is then nothing but the modular group of the torus
 \cite{four}. In this Letter we make manifest the unbroken N=1 supersymmetry 
of the spacetime-filling D3-brane action with dilaton-axion coupling, and 
investigate its other relevant symmetries in flat N=1 superspace.

\section{N=1 BI action in superspace}

In this section we briefly describe the N=1 BI action is superspace, which is 
the pre-requisite to our investigation in sect.~3. The BI action in Minkowski 
spacetime of signature $\h={\rm diag}(+,-,-,-,)$ is \cite{bi}
$$ S\low{\rm BI} = \fracmm{1}{\k^2}\int d^4x\,
\sqrt{ -\det (\h_{mn}+\k F_{mn})}~,\eqno(1)$$
where $F_{mn}=\pa_{m}A_{n}-\pa_{n}A_{m}$, $m,n=0,1,2,3$, 
and $\k$ is the dimensional coupling constant ($\k=2\p\a'$ in string theory).
The N=1 supersymmetric extension of the action (1) can be interpreted as 
the Goldstone-Maxwell action associated with partial (1/2) spontaneous 
supersymmetry breaking, N=2 to N=1, whose Goldstone fermion is photino of 
a Maxwell (vector) N=1 multiplet with respect to unbroken N=1 supersymmetry
\cite{two,two1}. Manifest supersymmetry does not respect the standard form (1)
 of the BI action. The complex bosonic variable, having the most natural 
supersymmetric extention, is given by 
$$ \o=\a +i\b~,\qquad \a = \fracmm{1}{4}F^{mn}F_{mn}~,
\quad \b = \fracmm{1}{4}F^{mn}\tilde{F}_{mn}~,\quad 
\tilde{F}^{mn}=\ha\ve^{mnpq}F_{pq}~.\eqno(2)$$
The BI Lagrangian (1) can be rewritten in terms of $\o$ and $\bar{\o}$  as
$$   \Lag\low{\rm BI}(\o,\bar{\o})= \Lag\low{\rm ~free}+ \Lag\low{\rm 
~int.} \equiv  -\,\fracm{1}{2}\left(\o+\bar{\o}\right)
+\k^2\o\bar{\o}\cy(\o,\bar{\o})~,
\eqno(3)$$
where the particular structure function $\cy(\o,\bar{\o})$ has been introduced,
  $$\cy(\o,\bar{\o}) = \fracmm{1}{1+\fracmm{\k^2}{2}(\o+\bar{\o})+
\sqrt{ 1+ \k^2(\o+\bar{\o})+\fracmm{\k^4}{4}(\o-\bar{\o})^2}}~~.
\eqno(4)$$
A supersymmetrization of the bosonic BI theory (1) in the form (3) amounts to 
replacing the field strength $F_{mn}$ by the N=1 chiral spinor superfield 
strength $W_{\a}$, and $\o$ by the N=1 chiral scalar superfield 
$K=\frac{1}{8}\bar{D}^2\bar{W}^2$, {\it viz.}  
$$ S\low{\rm sBI} = \frac{1}{4}\left(\int d^4xd^2\q\,W^2+{\rm h.c.}\right) 
+ \frac{\k^2}{8}\int d^4xd^4\q\,W^2\bar{W}^2\,\cy(K,\bar{K}) \eqno(5)$$ 
with {\it the same} structure function (4), so that the bosonic terms of 
eq.~(5) exactly reproduce eq.~(1). We use the standard notation,
$W^2= W^{\a}W_{\a}$ and $\bar{W}^2=\bar{W}_{\dt{\a}}\bar{W}^{\dt{\a}}$, and 
similarly for the N=1 flat superspace covariant derivatives $D^{\a}$ and
 $\bar{D}_{\dt{\a}}$ with $\a=1,2$ and $\dt{\a}=\dt{1},\dt{2}$. The gauge 
superfield strength $W_{\a}$ obeys the superfield Bianchi identities
$$ \bar{D}_{\dt{\a}}W\low{\a}=0 \quad{\rm and} \quad
 D^{\a}W_{\a}=\bar{D}_{\dt{\a}}\bar{W}^{\dt{\a}}~.\eqno(6)$$ 
In the chiral basis the gauge superfield strength reads
$$ W_{\a}(x,\q)=-i\j_{\a}(x) +\left[ \d\du{\a}{\b}D(x)
-i(\s^{mn})\du{\a}{\b}F_{mn}(x)\right]\q_{\b} 
+\q^2(\s^m\pa_m)_{\a\dt{\b}}\bar{\j}^{\dt{\b}}(x)~,\eqno(7)$$
where $\j_{\a}(x)$ is the fermionic superpartner (photino) of the abelian BI 
vector field $A_{m}$, and $D$ is the real auxiliary field. In the N=1 
super-BI theory (5) setting $D=0$ is consistent with its equations of motion
(this is called the `auxiliary freedom' \cite{jim1}).

The action (5) can be put into the simple `non-linear sigma-model' form 
\cite{two,two1} 
$$  S_{\rm sBI}=\int d^4xd^2\q\,X +{\rm h.c.},\eqno(8)$$
whose chiral superfield Lagrangian $X$ is determined via the recursive 
relation \cite{two,two1}
$$ X + \frac{\k^2}{4} X\bar{D}^2\bar{X} = \frac{1}{4}W^{\a}W_{\a}~~.
\eqno(9)$$

The BI action (1) is well-known to be invariant under non-trivial 
electric-magnetic duality \cite{schr}. This means that treating $F$ as a 
generic two-form, enforcing the Bianchi identity, $dF=0$, by means of a 
Lagrange multiplier (= dual vector potential) in the first-order action, and 
integrating out $F$ in favor of the Lagrange multiplier yield the dual action 
having {\it the same} form as eq.~(1) in terms of the dual vector potential. 
The same is true in N=1 superspace for the action (5) when introducing the 
dual N=1 superfield strength as an N=1 Lagrange multiplier, and integrating
over $W$ in the corresponding first-order action, i.e. after the N=1 
superfield Legendre transform \cite{two1}.  

Another highly non-trivial property of eq.~(5) is its invariance under the 
(non-linearly realized and spontaneously broken) second N=1 supersymmetry 
with rigid spinor parameter $\h_{\a}$ \cite{two},
$$\d_{\h} W_{\a}=\frac{1}{\k}\h_{\a} +\frac{\k}{4}\bar{D}^2\bar{X}\h_{\a}
+i\k(\s^{m}\bar{\h})_{\a}\pa_{m}X~.\eqno(10)$$ 
The transformations (10) are consistent with the N=1 Bianchi identities (6), 
and they realize a supersymmetry algebra. The invariance of the action (5) 
under the transformations (10) follows from the remarkable fact that 
$\int d^2\q\,\d_{\h}X=\frac{1}{2\k}\int d^2\q\,W^{\a}\h_{\a}$ is a total 
derivative in spacetime. 

To make manifest the hidden second supersymmetry of the the N=1 BI theory, one
can reformulate it in the formalism of non-linear realizations \cite{bik}. 
The Goldstone superfield $\J$ having the standard transformation law in the 
chiral version of the non-linearly realized supersymmetry \cite{sw}, 
$\d_{\h}\J=\frac{1}{\k}\h-2i\k(\J\s^m\bar{\h})\pa_m\J$, is given by 
$$ \J_{\a} = \fracmm{W_{\a}}{1+\frac{\k^2}{4}\bar{D}^2\bar{X}}+
 ~\ldots~,\eqno(11) $$
where the dots stand for the higher-order fermionic terms \cite{bik}. The new
Goldstone superfield $\J$ obeys the non-linear N=1 superspace constraints 
$$ \bar{\cd}_{\dt{\a}}\J\low{\a}=\cd\low{\a}\bar{\J}_{\dt{\a}}=0 \eqno(12)$$
that are also covariant under the second non-linearly realized supersymmetry. 
 The N=2 covariant derivatives in N=1 superspace \cite{two}
$$\cd_{\a}=D_{\a}+i\k^2(D_{\a}\J\s^m\bar{\J}+D_{\a}\bar{\J}\tilde{\s}^m\J)D_m
\quad{\rm and}\quad D_m=(\o^{-1})\du{m}{n}\pa_n~,\eqno(13)$$
where $\o\du{m}{n}=\d\du{m}{n}-i\k^2(\pa_m\J\s^n\bar{\J} 
+\pa_m\bar{\J}\tilde{\s}^n\J)$, form a closed algebra. The action (5) may be 
rewritten in terms of $\J$ and the N=2 covariant derivatives (13) as 
$$ S_{\rm sBI}= \frac{1}{4}\int d^4xd^2\q\,{\cal E}^{-1}\J^2 + 
{\rm h.c.}~,\eqno(14)$$
whose N=1 chiral superfield  ${\cal E}^{-1}=1+\frac{\k^4}{4}
\bar{D}^2\bar{X}+\ldots$, should transform as a density under the second 
supersymmetry, $\d_{\h}{\cal E}^{-1}=-2i\k\pa_m({\cal E}^{-1}\J\s^m\bar{\h})$.

Both the electric-magnetic self-duality and the second non-linearly realized 
supersymmetry of the N=1 BI action may have been expected from its anticipated 
connection to the D3-brane action. It is just these key properties that allow 
one to identify the N=1 BI action with the low-energy effective action of the 
spacetime-filling D3-brane in the case of slowly varying fields. Any direct 
gauge-fixing of the kappa-symmetric D3-brane action \cite{one} would yield 
highly involved supersymmetry transformations, whose precise relation to the 
standard N=1 superspace transformations implies complicated field 
redefinitions. We didn't attempt to establish this connection explicitly.

\section{N=1 BI action with dilaton-axion coupling}

The bosonic BI action coupled to a background dilaton $\f$ and axion $C$ reads
$$ S_{\rm bosonic} =\fracmm{1}{4\p} \int d^4x\,\sqrt{ -\det (\h_{mn}+e^{-\f/2}
 F_{mn})}+\fracmm{1}{32\p}i\ve^{mnpq}CF_{mn}F_{pq}~.\eqno(15)$$
The dilaton-axion background now plays the role of the effective coupling
constant, so that we chose $\k=1$ for simplicity. We also rescaled the BI 
action by a factor of $4\p$, in order to make it invariant under the T-duality
 transformations, $C\to C + n$, where $n\in {\bf Z}$, because $C$ multiplies 
the topological density in eq.~(15).

It is not difficult to supersymmetrize eq.~(15) in N=1 superspace, by using
the results of sect.~2. First, let's define a complex scalar
$$ \r= e^{-\f} +iC~,\eqno(16)$$
and assume that it belongs to an N=1 chiral superfield,
$$ \F =\r +\q^{\a}\l_{\a}+\q^2F~,\eqno(17)$$
where we have introduced the physical dilatino $\l_{\a}$ and the `auxiliary' 
field $F$. This is not quite innocent procedure in the theories with higher
derivatives, because the field $F$ should be truly auxiliary or, at least, 
$F=0$ should be a solution to the equations of motion (the auxiliary freedom).
Equation (5) implies the N=1 supersymmetric extension of eq.~(15) in the form
$$\eqalign{
4\p S ~&~ =~  \frac{1}{4}\left(\int d^4xd^2\q\,\F W^2+{\rm h.c.}\right) \cr 
~&~ + \frac{1}{32}\int d^4xd^4\q\,(\F+\bar{\F})^2W^2\bar{W}^2\,
\cy\left(\ha(\F+\bar{\F})K,\ha(\F+\bar{\F})\bar{K}\right)~,\cr}\eqno(18)$$
with {\it the same} function $\cy$ defined by eq.~(4) at $\k=1$, where we
have used the identity
$$ D^2W^2-\bar{D}^2\bar{W}^2=i\ve^{mnpq}F_{mn}F_{pq}~.\eqno(19)$$

The N=1 Legendre transform of the action (18) with respect to the gauge
superfield $W$ yields the dual N=1 superspace action that has {\it the same\/}
form (18) in terms of the dual N=1 superfield strength {\it and\/} the dual 
coupling 
$$ \tilde{\F} =\fracmm{1}{\F}~~.\eqno(20)$$
Together with imaginary shifts of $\F$ by integers the S-duality transformation
 (20) generates the full $SL(2,{\bf Z})$ duality, as required. In fact, the 
action (18) is invariant under the continuous $SL(2,{\bf R})$ duality, as it 
belongs to the class of the $SL(2,{\bf R})$ duality invariant actions 
constructed in ref.~\cite{kuz}. Of course, in quantum theory only 
$SL(2,{\bf Z})$ survives. 

The $SL(2,{\bf R})$ duality invariant dilaton and axion kinetic terms to be
added to eq.~(18),
$$\Lag(\f,C)=\frac{1}{2}(\pa_m\f)^2+\frac{1}{2}e^{2\f}(\pa_m C)^2~,\eqno(21)$$
are given by the K\"ahler non-linear sigma-model with a K\"ahler potential
$$ K(S,\bar{S})=-\ln (S+\bar{S})~.\eqno(22)$$ 
The N=1 supersymmetrization of eq.~(21) in superspace is straightforward,
$$ S_{\rm kin.}=-\int d^4x d^4\q\,\ln (S+\bar{S})~~.\eqno(23)$$

There is, however, a problem with another (non-linearly realized) 
supersymmetry. A variation of the leading terms in eq.~(18) yields 
$$ \d_{\h}\Lag =\ha \int d^2\q\,\F W^{\a}\h_{\a} +{\rm h.c.}~,\eqno(24)$$
which is a total derivative only for a {\it constant\/} dilaton-axion 
background $\F$. Yet another problem is the auxiliary freedom of $F$.  

The way out of both problems may be the assignment of dilaton and axion to an
N=1 {\it linear\/} multiplet $G$, instead of the N=1 chiral multiplet $\F$. 
As regards the bosonic action (15), this means trading $C$ against a gauge 
two-form $B$, at the expense of giving up the manifest $U(1)$ gauge invariance,
{\it viz.}
$$ \int CF\wedge F=-\int dC\wedge(A\wedge F)=\int {}^*dB\wedge \Q~,\eqno(25)$$
where the star denotes the Poincar\'e dual, ${}^*(dC)=dB$ and $\Q=A\wedge F$ 
is the abelian Chern-Simons three-form. In N=1 superspace a real linear 
supefield $G$ is defined by the constraints
$$ D^2G=\bar{D}^2G=0~.\eqno(26)$$
It consists of a real scalar (dilaton), an antisymmetric tensor $(B)$ subject
to the gauge transformation $\d B=d\x$ with the one-form gauge parameter $\x$,
a dilatino $\l_{\a}$, and {\it no\/} auxiliary fields. The two-form $B$ enters
 the superfield $G$ only via its field strength $dB$.

The leading term in eq.~(18) can then be rewritten to the form
$$\frac{1}{4}\left(\int d^4xd^2\q\,\F W^2+{\rm h.c.}\right)=
 \frac{1}{4}\int d^4xd^4\q\,(\F+\bar{\F})\O~,\eqno(27)$$
where we have introduced the Chern-Simons superfield $\O$ via the equations
$$ W^2=\frac{1}{2}\bar{D}^2\O~,\quad \bar{W}^2=\frac{1}{2}D^2\O~.\eqno(28)$$
By using a solution $W_{\a}=-\frac{1}{4}\bar{D}^2D_{\a}V$ to the Bianchi 
identities (6), in terms of the real gauge scalar superfield $V$ subject to 
the gauge transformations $V\to V +i(\L-\bar{\L})$, with 
$\bar{D}_{\dt{\a}}\L=0$, we easily 
find $\O=-\frac{1}{4}(D^{\a}V)W_{\a}+{\rm h.c.}~$

The full action given by a sum of eqs.~(18) and (23) is now dependent upon the
 chiral superfields $\F$ and $\bar{\F}$ only through their linear combination 
$\ha(\F+\bar{\F})$, so that it is 
possible to dualize this action in terms of the linear superfield $G$ by 
Legendre transform.~\footnote{The possibility of such transformation was 
noticed in ref.~\cite{kuz}.} We replace in eqs.~(18) and (23) the combination 
$\ha(\F+\bar{\F})$ by a general real superfield $U$, and add extra term
$$ \int d^4xd^4\q\,UG \eqno(29)$$
to the action (18). On the one hand side, varying eq.~(26) with respect to $G$ 
(in fact, with respect to a potential $J_{\a}$ in the general solution 
$G=D^{\a}\bar{D}^2J_{\a}+\bar{D}_{\dt{\a}}D^2\bar{J}^{\dt{\a}}$ to the 
defining constraints (26)), we get $U=\ha(\F+\bar{\F})$ back. On the other hand
side, varying with respect to $U$ in the action
$$ S = \int d^4xd^4\q\,\left[- \ln U + UG+ \frac{1}{8\p}U\O   
+ \frac{1}{32\p}U^2W^2\bar{W}^2\,\cy(UK,U\bar{K})\right]\eqno(30)$$
we find an algebraic equation on $U$:
$$ \fracmm{1}{U} = 
\left(G+\fracmm{1}{8\p}\O\right)+\fracmm{1}{32\p}W^2\bar{W}^2
\left(2U\cy(UK,U\bar{K}) +U^2\fracmm{\pa\cy(UK,U\bar{K})}{\pa U}\right)~~.
\eqno(31)$$
Since $W_{\a}W_{\b}W_{\g}=0$ due to the anti-commutativity of $W_{\a}$, the
second term on the right-hand-side of recursive relation (31) can be 
considered as an `exact' perturbation. This leads to a complete solution to 
eq.~(31) in the form
$$ U^{-1}=G_{\rm mod}+\fracmm{1}{32\p}W^2\bar{W}^2
\left(\fracmm{2\cy(G^{-1}_{\rm mod}K,G^{-1}_{\rm mod}\bar{K})}{G_{\rm 
mod}} -\fracmm{\pa\cy(G^{-1}_{\rm mod}K,G^{-1}_{\rm mod}\bar{K})}{\pa
G_{\rm mod} }\right)~~,\eqno(32)$$
where we have introduced the `modified' N=1 linear multiplet $G_{\rm mod}$ as
$$ G_{\rm mod} = G +\fracmm{1}{8\p}\O~.\eqno(33)$$ 
The appearance of the N=1 Chern-Simons superfield $\O$ is quite
natural from the point of view of string theory and D-branes, where 
Chern-Simons-type couplings (in components) are known to appear in the 
famous Green-Schwarz anomaly cancellation mechanism and in the (dual) D-brane
actions. In particular, the dilaton superfield $G$ must transform under the
$U(1)$ gauge transformations as  
$$ \d G = \fracmm{i}{32\p}(D^{\a}\L)W_{\a} + {\rm h.c.}\eqno(34)$$ 
in order to make $G_{\rm mod}$ gauge-invariant. Equations (26) and (28) lead to
 the manifestly gauge-invariant constraints on $G_{\rm mod}$,
$$ \bar{D}^2G_{\rm mod}=\fracmm{1}{4\p}W^2~,\qquad  
 D^2G_{\rm mod}=\fracmm{1}{4\p}\bar{W}^2~.\eqno(35)$$
Such couplings were extensively studied in superspace (see e.g., 
ref.~\cite{mar} for a recent review), while the relevant superspace geometry
appears to be closely related to a three-form N=1 multiplet introduced in
ref.~\cite{ga}. 

Substituting the solution (32) into the action (30) yields the dual action in 
the form
$$ S = \int d^4xd^4\q\,\left\{ \ln G_{\rm mod}    
+ \frac{1}{32\p}W^2\bar{W}^2G^{-2}_{\rm mod}\cy(G^{-1}_{\rm mod}K,G^{-1}_{\rm 
mod}\bar{K}) \right\}~~.\eqno(36)$$
By construction this action is equivalent (dual) to the action given by a sum
of eqs.~(18) and (23). However, it seems to be much easier to find a 
supersymmetric completion of the action (36) with respect to the second 
(spontaneously broken) supersymmetry, since the action (36) is given by the 
{\it full} N=1 superspace integral, while the constraints (35) are also easy
to be covariantized. The second non-linearly realized supersymmetry with the 
transformation law $\d W_{\a}=\h_{\a} +\ldots$ implies a non-trivial 
transformation law of $G_{\rm mod}$ as well, because of the constraint (35),
$$ \d_{\h} G_{\rm mod}=-\frac{1}{8\p}(\h^{\a}D\low{\a}V
+\bar{\h}_{\dt{\a}}\bar{D}^{\dt{\a}}V)+\ldots~.\eqno(37)$$
The minimal, manifestly N=2 covariant version of the constraints (35)
 is given by
$$ \bar{\cd}^2\tilde{G}_{\rm mod}=\fracmm{1}{4\p}\J^2~,\qquad  
 \cd^2\tilde{G}_{\rm mod}=\fracmm{1}{4\p}\bar{\J}^2~,\eqno(38)$$
where we have substituted the Maxwell-Goldstone N=1 superfield $W$ by the N=1 
Goldstone superfield $\J$, and the N=1 linear (dilaton-axion) superfield 
$G_{\rm mod}$ by its fully covariant counterpart $\tilde{G}_{\rm mod}$. 
The superfield $W$ obeys the `canonical' constraints (6) but it has the 
complicated transformation law (10), whereas  the N=1 Goldstone superfield 
$\J$ has the `canonical' transformation law under the second supersymmetry
but it obeys the complicated constraints (12). The same remarks also apply 
to $G_{\rm mod}$ and $\tilde{G}_{\rm mod}$, respectively. 

The defining constraints (38) on $\tilde{G}_{\rm mod}$ are consistent with the
 constraints (12) because of the identities 
$$\cd\low{\a}\cd\low{\b}\cd\low{\g}=
\bar{\cd}_{\dt{\a}}\bar{\cd}_{\dt{\b}}\bar{\cd}_{\dt{\g}}=0\eqno(39)$$ 
that follow from the definitions (13). The fully covariant action is thus of 
the form
$$ S = \int d^4x d^4\q\,E^{-1}\ln\tilde{G}_{\rm mod}~,\eqno(40)$$
where we have introduced a density $E^{-1}(\J,\bar{\J},\tilde{G}_{\rm mod})$ 
in the full N=1 superspace. 

\section*{Acknowledgements}

The author would like to thank the Institute for Theoretical Physics of the
University in Hannover, and the Department of Physics of the University in
Kaiserslautern, Germany, for kind hospitality extended to him during a 
preparation of this paper. This work was supported by the German Science 
Foundation (DFG) under the Federal Research Programm `String Theory' and the 
Volkswagen Grant from the University of Kaiserslautern.

\end{document}